\documentclass[epj,nopacs,final]{svjour}
\usepackage{graphics}
\usepackage{latexsym}
\usepackage{cite}
\usepackage{subfigure,wrapfig,multirow}
\usepackage{epsfig,color,rotating,amsmath,delarray,array}
\usepackage{makeidx,pifont,float,amssymb}
\definecolor{gray01}{gray}{0.9}
\definecolor{gray02}{gray}{0.8}
\definecolor{gray03}{gray}{0.7}
\definecolor{gray04}{gray}{0.6}
\definecolor{gray05}{gray}{0.5}
\definecolor{gray06}{gray}{0.4}
\definecolor{gray07}{gray}{0.3}
\definecolor{gray08}{gray}{0.2}
\definecolor{gray09}{gray}{0.1}

\newcommand{\bc}{\begin{center}}
\newcommand{\ec}{\end{center}}
\newcommand{\be}{\begin{equation}}
\newcommand{\ee}{\end{equation}}

\begin{document}

\title{Light-quark baryons}
\titlerunning{Light-quark baryons}
\authorrunning{V. Burkert, E. Klempt and U. Thoma}
\author{Volker Burkert$^{1}$, Eberhard Klempt$^{2}$ and Ulrike Thoma$^{2}$}

\institute{$^{1}$Thomas Jefferson National Accelerator Facility, 12000 Jefferson Avenue, 
Newport News, Virginia 23606, USA \label{addr1a}\\
$^{2}$Helmholtz-Institut f\"ur Strahlen- und Kernphysik,
Universit\"at Bonn, Germany\label{addr2a}}

\abstract{This is a contribution to the review ``50 Years of Quantum Chromdynamics"
edited by F. Gross and E. Klempt, to be published in EPJC. The contribution reviews the
new baryon resonances derived from photoproduction experiments. Implications of the new
results for the interpretation of baryons are discussed. }
\date{Received: \today / Revised version:}

\mail{klempt@hiskp.uni-bonn.de}

\maketitle
\section{Why $N^*$'s\,?}
This was the question with which Nathan Isgur opened his talk at
$N^*$2000~\cite{Isgur:2000ad} held at the Thomas Jefferson National Accelerator Facility
in Newport News, VA, one year before he passed away, much too early.
 He gave three answers:

First, nucleons are the stuff of which our world is made. 
The $N^*$'s and $\Delta^*$'s are of great importance in the development of the Universe, when
hadrons materialized from a soup  of quarks and gluons at some 
10\,$\mu$s after the big bang.
The full spectrum of excited baryon states including those carrying strange\-ness must be included in hadron gas models that simulate the freeze-out behavior observed in hot-QCD calculations. These simulations aim at finding the underlying processes, to pin-point the "critical point" of the phase transition that is expected to occur between the QGP phase and the hadron phase at a temperature near 155~MeV. Experiments are ongoing at CERN, RHIC and planned at FAIR to study the phase diagram of strongly interacting matter, e.g. by varying the collision energy.

Second, nucleons are the simplest system in which the non-abelian
character of QCD is manifest. The proton consists of three (constituent)
quarks since the number of colors is three.

Third, baryons are sufficiently complex to reveal physics to us hidden in the mesons.
Gell-Mann and Zweig did not develop their
quark model along mesons, their simple structure allowed for different
interpretations. {\it Three} quarks resulted in a baryon structure 
that gave - within SU(3) symmetry - the octet and the decuplet containing the famous $\Omega^-$.

Isgur made many important contributions to the development of the quark model.
With Karl he developed the idea that gluon-mediated interactions between quarks
bind them into hadrons and constructed a quark model of  baryons~\cite{Isgur:1977ef}. This
was a non-relativistic model, hardly justifiable. With Capstick he relativized
the model~\cite{Capstick:1986ter}, but surprisingly, the pattern of predicted
resonances remained rather similar. Isgur always defended the basic principles: 
hadrons have to be understood
in terms of constituent quarks bound in a confining potential and additionally
interacting via the exchange of ``effective" gluons.

Nearly 20 years later, Mei\ss ner ended his contribution~\cite{Meissner:2019cpk}
to the $N^*$2019
conference held in Bonn, Germany, by stating: ``Forget the quark model". 
We need to ask: What has happened in these two decades? What did we
know before? What have we learned?

Mapping the excitation 
spectrum of the nucleon (protons and neutrons) 
and understanding the effective degrees of freedom are important and  most challenging tasks of hadron physics. 
A quantitative description of the spectrum and properties of excited nucleons 
must eventually involve solving QCD for a complex strongly interacting 
multi-particle system. 
The experimental $N^*$ program currently 
focuses on the search for new excited states in the mass range just below
and above 2\,GeV 
using energy-tagged photon beams in the few GeV range, and on the study 
of resonances, their properties, and their internal structure, e.g. in cascade decays
and in meson electro-production.  

\section{$N^*$'s: how?}
The expected spectrum of nucleon and $\Delta$ excitations if rather rich. Even
in the lowest excitation mode with $l_\rho=1$ or $l_\lambda=1$, we expect five
$N^*$ and two $\Delta^*$ states; they are all well established. But already in the second
excitation mode, the quark model predicts 13  $N^*$ and 8 $\Delta^*$ states The resonances
have quantum numbers $J^P=1/2^+, \cdots , 7/2^+$ and isospin $I=1/2$ or $3/2$, respectively. All these
21 resonances are expected to fall into a mass range of, let's say, 1600 - 2100\,MeV.
This complexity of the light-quark ($u$ \& $d$ quarks) baryon 
excitation spectrum complicates the experimental search for individual states, especially 
since, as a result of the strong interaction, 
these states are broad, the typical width being 150-300\,MeV.  They overlap, interfere,
and often several resonances show up in the same partial wave.
Compared to $\pi\pi$ scattering experiments, additional complications due to
the nucleon spin emerge in $\pi N$ elastic scattering. Now there are two complex 
amplitudes to be determined, for spin-flip and spin-non-flip scattering.

Pion scattering off nucleons was mostly performed in the pre-QCD era.
Nearly all excited nucleon states listed in the Review 
of Particle Physics (RPP) prior to 2012 have been observed in elastic pion scattering 
$\pi N \to \pi N$.  However there are important limitations in the sensitivity to the 
higher-mass nucleon states. These may have very small $\Gamma_{\pi N}$ decay widths, and   
their identification becomes exceedingly difficult in elastic scattering.
Three groups extracted the real and imaginary parts
of the $\pi N$ partial-wave amplitude from the data \cite{Hohler:1979yr,Cutkosky:1979fy,Arndt:2006bf}. 
Their results are still used as 
constraints in all modern analyses of
photo-induced reactions. 

\begin{figure*}[pt]
\begin{tabular}{cc}
\hspace{-2.mm}\includegraphics[height=.231\textwidth]{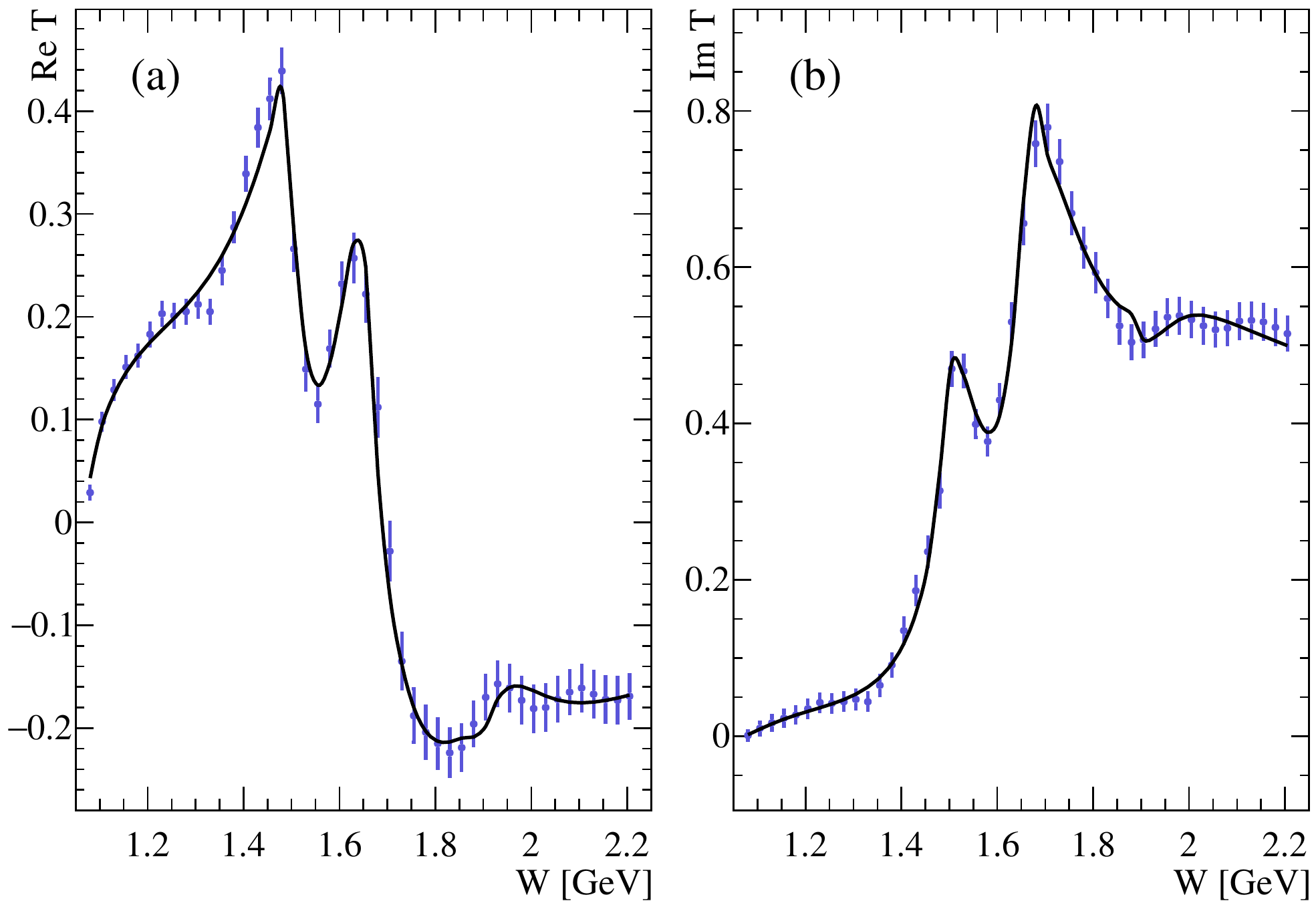} \hspace{-1mm}
\includegraphics[height=.234\textwidth]{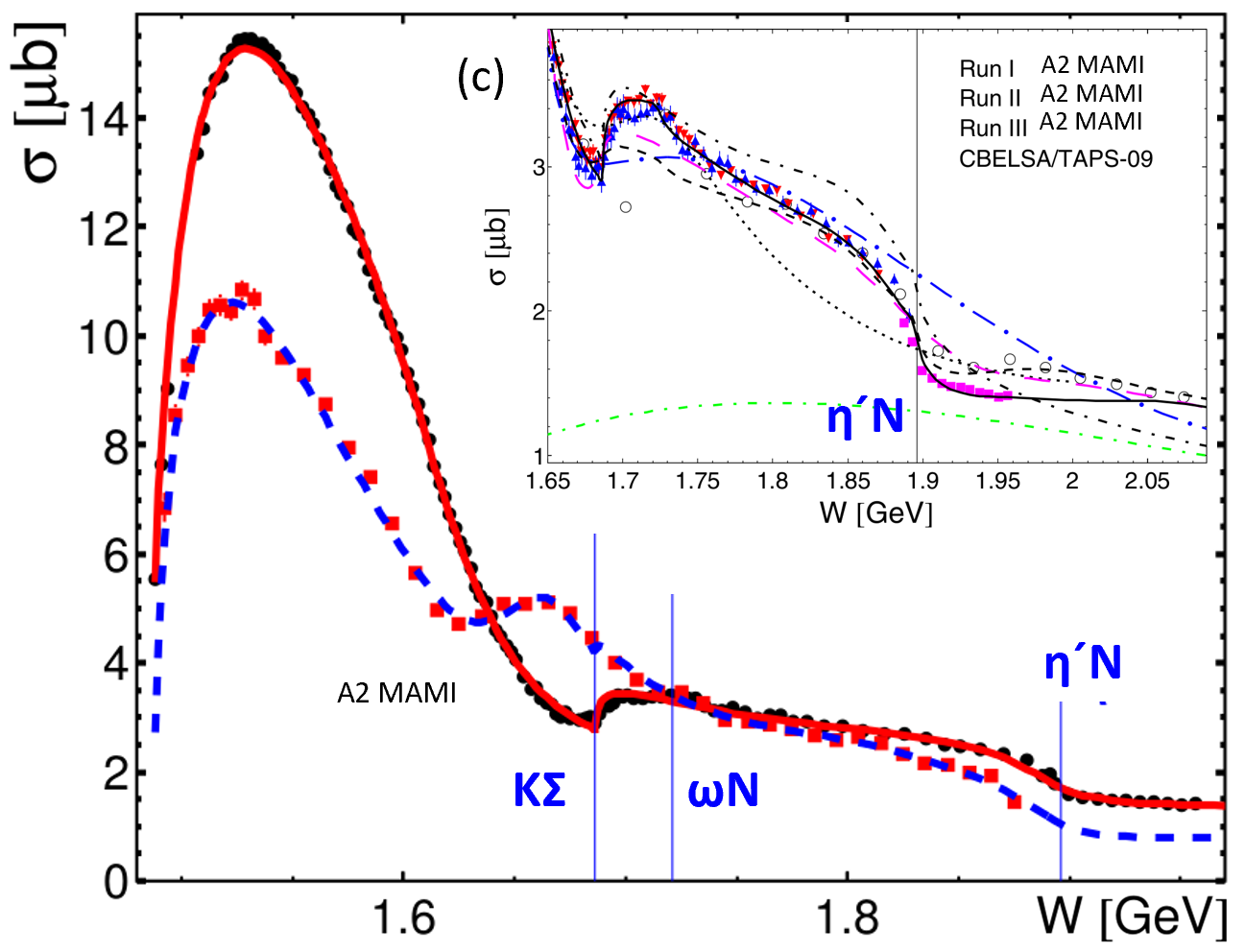}\hspace{-1.mm}
\includegraphics[height=.234\textwidth]{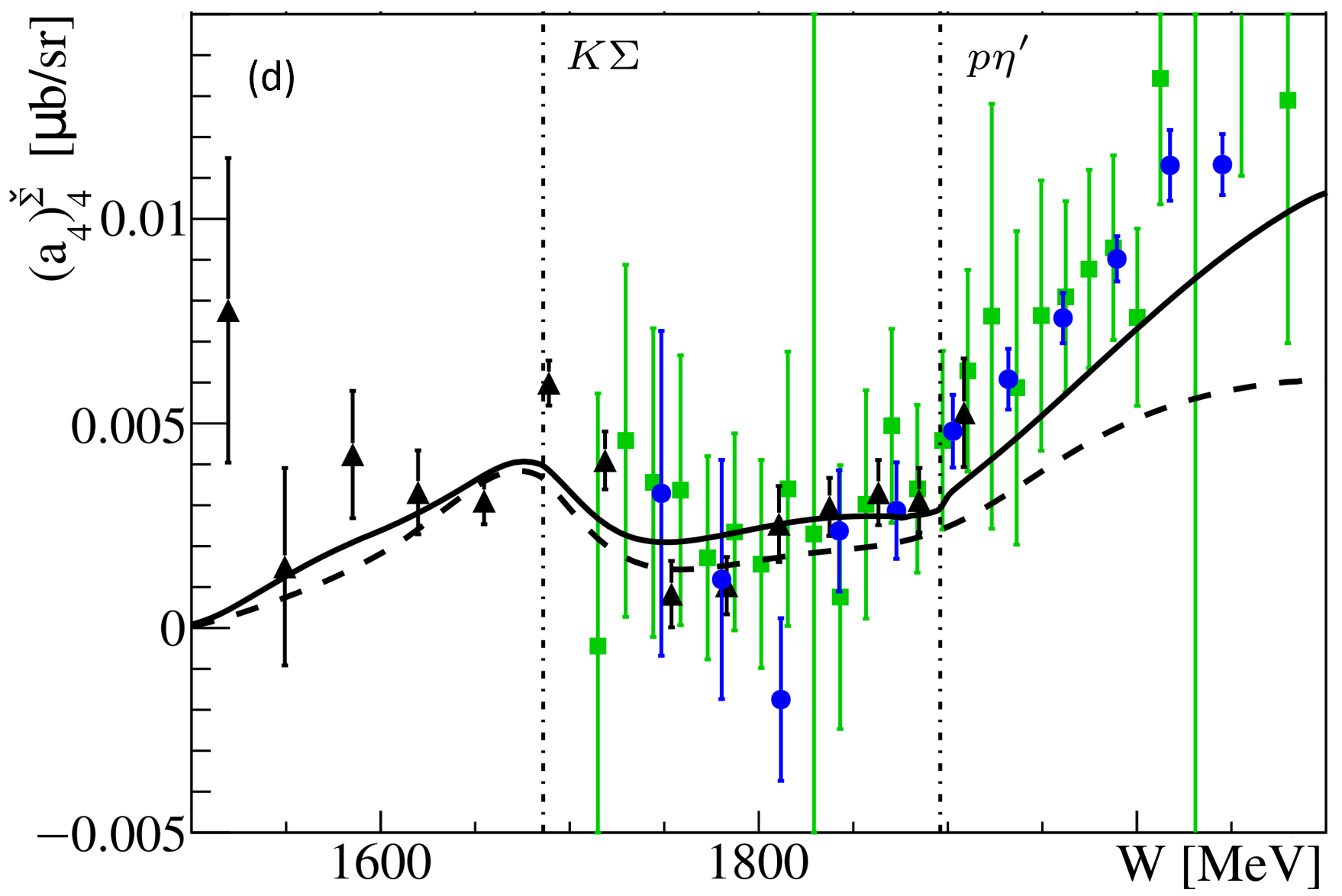}
\end{tabular}
\caption{\label{fig:S11}(a),(b): Real and imaginary part of the $S_{11}$ $\pi N$ scattering amplitude.
Resonances in this partial wave have quantum numbers $J^P=1/2^-$.
Clearly seen are $N(1535)1/2^-$ and $N(1650)1/2^-$.
There is no convincing evidence for any resonance above 1700\,MeV. Data points are
from \cite{Hohler:1979yr}, errors are estimates, the curve represents a recent Bonn-Gatchina (BnGa) fit. (c): Total cross sections for $\gamma p\to \eta\,p$ and $\gamma n\to \eta\,n$. Important thresholds are marked by lines. The inset shows the $\eta^\prime$ threshold region for $\eta$-photoproduction off the proton (picture adapted from~\cite{Tiator:2018heh,A2:2017gwp}). (d): The Legendre coefficient of the polarization observable $\Sigma$
$(a_4)_4 ^\Sigma$ exhibits a cusp at the $\eta^\prime$ threshold \cite{CBELSATAPS:2020cwk}. The data stems from GRAAL (black), CBELSA/TAPS (blue) and CLAS (green) Picture taken from~\cite{CBELSATAPS:2020cwk}. (c),(d): see publications\cite{Tiator:2018heh,A2:2017gwp,CBELSATAPS:2019ylw} for references to the data. 
}
\end{figure*}

\begin{figure}[pt]
\begin{tabular}{cc}
\includegraphics[width=0.95\columnwidth]{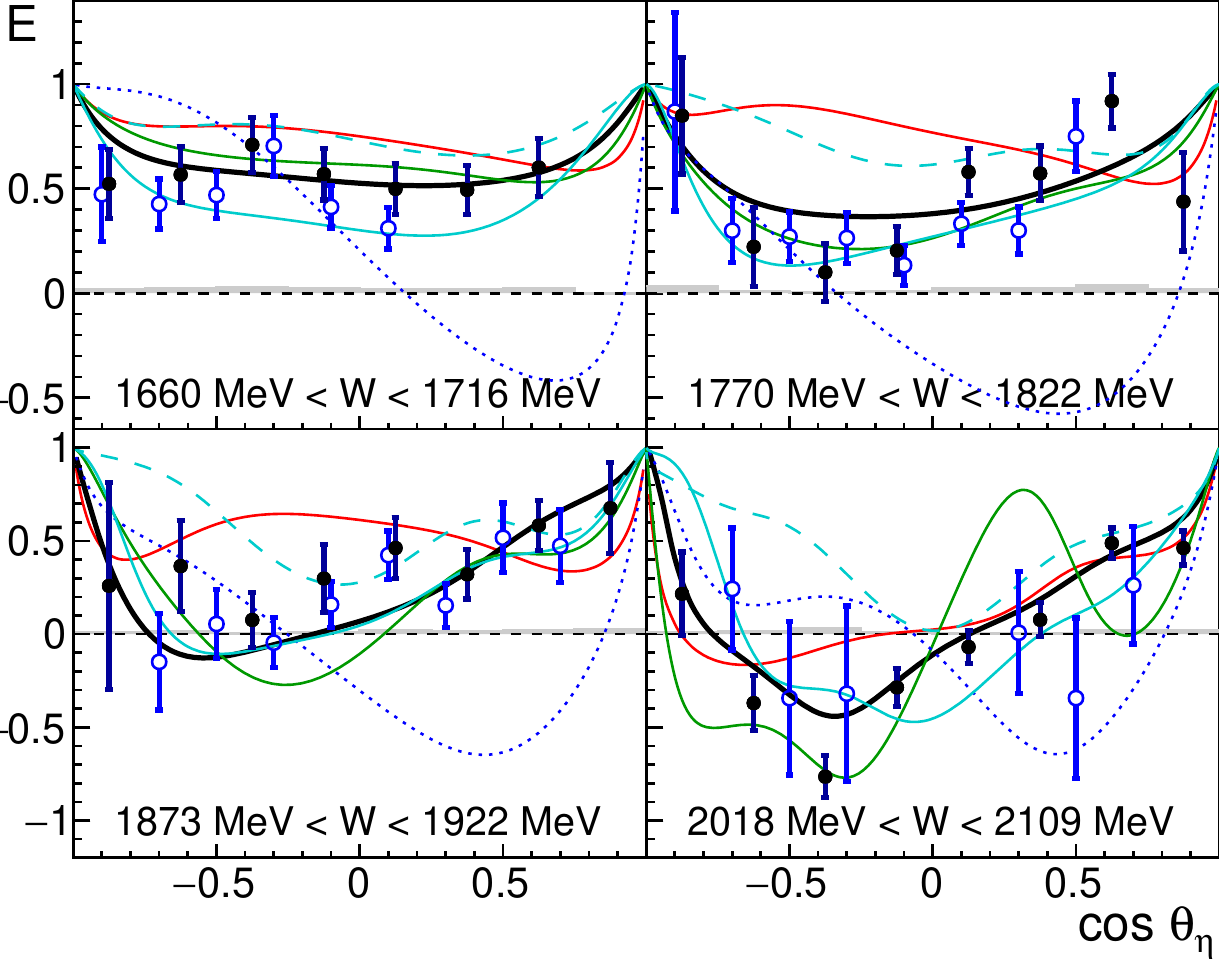}
\end{tabular}
\caption{\label{fig:eta-E-cb}
The double polarization observable E as 
a function of $\cos{\theta_\eta}$ in the 
cms for selected energy bins, black: CBELSA/TAPS~\cite{CBELSATAPS:2019ylw}, 
blue: CLAS data~\cite{CLAS:2015pjm} (due to different binning,
the energies differ by up to half of the bin size).
Colored curves: Predictions from different PWAs (see publication for references), black: BnGa-fit including the 
data shown here and further new polarization data.  
Figure adapted from~\cite{CBELSATAPS:2019ylw}. }
\end{figure}

Figure~\ref{fig:S11}a,b shows the real and imaginary part of the $S_{11}$ amplitude for 
$\pi N$ scattering. The imaginary part peaks at 1500\,MeV and just below 1700\,MeV indicating the presence of two
resonances, $N(1535)1/2^-$ and $N(1650)1/2^-$.  These are known since long and established.
Above, there is no clearly visible sign for any additional resonance. Higher-mass resonances -- if they
exist -- must have very small $\Gamma_{\pi N}$ decay widths.

Estimates for alternative decay channels have 
been made in quark model calculations~\cite{Capstick:1993kb}.  This has
led to major experimental efforts at Jefferson Lab, ELSA and MAMI
to determine differential cross sections and (double) polarization observables 
for a variety of meson  photoproduction channels. Spring-8 at Sayo in Japan and  the
ESRF in Grenoble, France, made further 
contributions to the field.

Figure~\ref{fig:S11}c,d
shows an example. In Fig.~\ref{fig:S11}c, the total cross section for $\eta$ photoproduction
off protons and off neutrons is shown~\cite{Tiator:2018heh,A2:2017gwp}. 
They are dominated
by $N(1535)1/2^-$ $\to N\eta$ interfering with
$N(1650)1/2^-$. 
The opening of important channels is indicated by vertical lines. At the
$\eta^\prime$ threshold, the intensity suddenly drops: significant
intensity goes into the $N\eta^\prime$ channel. This is a strong argument
in favor of a resonance at or close to the $p\eta^\prime$ threshold. It also clearly    
demonstrates the advantage of investigating different final states and production mechanisms. 
In contrast to the $\pi N$-$S_{11}$ scattering amplitude, here, already in the total $\eta$-photoproduction cross section, a structure relating to $N(1895)1/2^-$ becomes  visible. Furthermore, in Fig.~\ref{fig:S11}d, the result of a fit with Legendre moments
to the so-called $\Sigma$ polarization observable for
$\gamma p\to \eta\,p$  is compared to two energy-dependent solutions
of the BnGa coupled-channel analysis. Plotted is the coefficient $(a_4)_4 ^\Sigma$
 of the Legendre expansion which receives (among others) a contribution
from the interference of the $S$-wave with the $G$-wave. Data from different
experiments are given with their error bars. The curves represent BnGa
fits with (solid curve) and without (dashed curve) inclusion of data on $\gamma p\to \eta^\prime\,p$.
The $N(2190)7/2^-$ ($G$-wave) was included in both fits. From 1750\,MeV to the $p\eta^\prime$-threshold the coefficient is approximately constant, then at the 
$p\eta^\prime$-threshold, the fit result shows an almost linear rise towards positive 
values. This change of the coefficient at about 1.9\,GeV indicates the presence of a cusp. 
The strong cusp is an effect of
the  $p\eta^\prime$ threshold [$E_\gamma = 1447$\,MeV (W = 1896\,MeV)], the $N\eta^\prime$ amplitude
must be strongly rising above threshold. Indeed, the inclusion of the full data
set on $\gamma p\to p\eta^\prime$ (cross sections, polarization
observables) into the BnGa data base had already confirmed 
the existence of a new $N(1895)1/2^-$ resonance with a significant coupling to $p\eta$ and $p\eta^\prime$~\cite{Anisovich:2017pox,Anisovich:2018yoo}, first observed in~\cite{Anisovich:2011fc}.

This resonance was not seen in classical analyses of $\pi N$ elastic
scattering data\footnote{H\"ohler and Manley had claimed a similar state that had been combined with Cutkovsky's result to $N(2090)$.}. The example shows the importance of inelastic channels and
of coupled-channel analyses. Thresholds can be identified by the missing intensity
in other channels, cusp effects can show up, all these effects need to be considered and finally contribute to 
find the correct solution. High-precison and high-statistics
data are required as well as a large body of different polarization data. 

\begin{figure*}[pt]
\centering
\includegraphics[width=.95\textwidth]{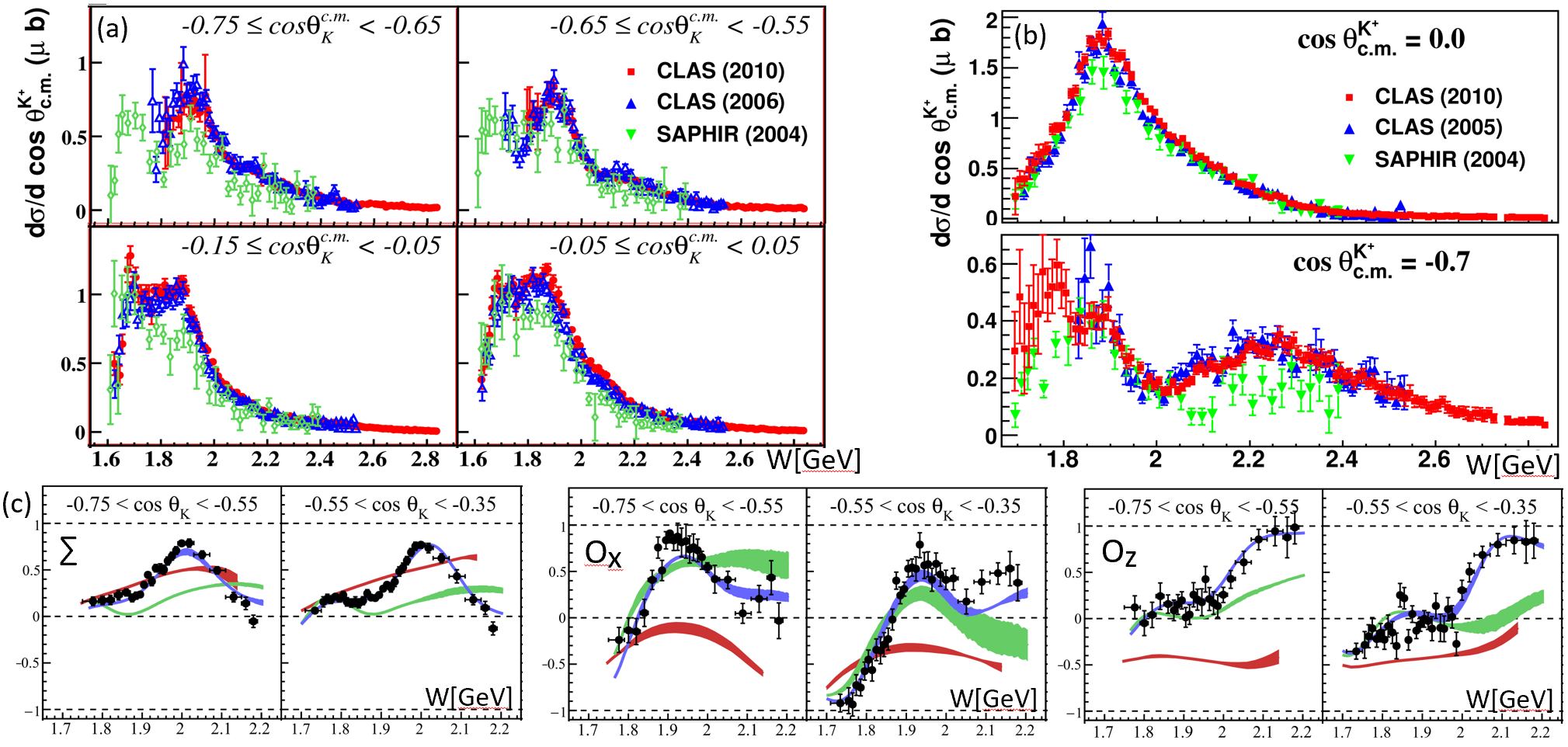}
\caption{Invariant mass dependence of the $\gamma p \to K^+\Lambda$~\cite{CLAS:2009rdi} (a) and $\gamma p \to K^+\Sigma$~\cite{CLAS:2010aen} (b) differential cross sections  for selected bins in the polar angle.
(c) Examples for polarization observables determined for $\gamma p \to K^+\Lambda$ (only selected bins shown)~\cite{CLAS:2016wrl}.
Curves: PWA-predictions from ANL-Osaka (red) and BnGa 2014 (green). Blue: BnGa 2014-refit including the data shown. 
(a)-(c): For references to the data and the PWAs see~\cite{CLAS:2009rdi,CLAS:2010aen,CLAS:2016wrl}, Picture adapted from~\cite{CLAS:2009rdi,CLAS:2010aen,CLAS:2016wrl} 
}
\label{KLambda-crs}
\end{figure*}

\section{Photoproduction of exclusive final states}

In the photoproduction of a single pseudoscalar meson like $\gamma p\to \eta\,p$,  not only
the proton has two spin states but also the photon has two possible spin orientations.
In electroproduction,
the virtual photon can also be polarized longitudinally. But even for experiments with real photons, there are
four complex amplitudes to be determined. There is a large number of
observables: the target nucleon can be polarized longitudinally, i.e. in
beam direction, or transversely, the photon can carry linear or circular polarization.
The final-state nucleon can carry polarization along its flight direction or
perpendicular to the scattering plane. There is an intense discussion in the literature on how many independent measurements
have to be performed to determine the four complex amplitudes, see Ref.~\cite{Thiel:2022xtb}. In practice, 
energy-independent analyses in bins of the invariant mass
were only done for the very low energy
region~\cite{Osmanovic:2019mux,Osmanovic:2021rck}
or with additional assumptions (see \cite{Anisovich:2017bsk,Hartmann:2014mya,Svarc:2021gcs}
and references therein). 

In most cases, energy-dependent analyses have been performed to extract the information hidden in the photoproduction data. Here 
the polarization data, in particular those with polarized photon beam and polarized target nucleons, were decisive to reduce ambiguities of the 
solutions. The double polarisation observable $E$ is one of the beam-target-observables; it requires a circularly polarized photon beam and a longitudinaly polarized target. Examples of $E$ for selected W-bins are shown in Fig.~\ref{fig:eta-E-cb} for $\gamma p \to p \eta$~\cite{CBELSATAPS:2019ylw}. The data are compared to the predictions of different PWA solutions (colored curves). The curves scatter over a wide range indicating the high sensitivity of the polarisation observable on differences in the contributing amplitudes. A new BnGa fit returned masses and widths of $N^*$-resonances and their $N\eta$-branching 
fractions~\cite{CBELSATAPS:2019ylw}, several of them unknown before. Interestingly  
a $N(1650)1/2^{-} \to N\eta$-branching fraction of $0.33\pm0.04$ was found while in the RPP'2010, a value of only $0.023\pm 0.022$ was given. Recently, also within the Jülich-Bonn dynamical coupled channel approach, a $N\eta$-residue for $N(1650)1/2^{-}$ was found, larger by almost a factor of two compared to earlier analyses, after inclusion of the new polarisation data~\cite{Ronchen:2022hqk}.    
Historically, the large $N(1535)1/2^{-} \to N\eta$ branching fraction and the small one for $N(1650)1/2^{-}$ $\to N\eta$ has played a significant role in the development of the quark model~\cite{Isgur:1978xj}, of theories based on coupled-channel chiral effective 
dynamics~\cite{Kaiser:1995cy} and led to several interesting interpretations of the 
low mass $1/2^{-}$-resonances (for references see~\cite{CBELSATAPS:2019ylw}). 
The old values from 2010 were obtained without the constraints provided by the new high quality (double) polarization data covering almost the complete solid angle. 
The impact of polarization observables on the convergence
of different PWA-solutions was e.g. also very clearly demonstrated
in a common study of pion-photoproduction~\cite{Anisovich:2016vzt}. 

In hyperon decays, the polarization of the $\Lambda$ or $\Sigma^\circ$ 
can be determined by analyzing the parity violating decay $\Lambda \to p \pi^-$. Thus the spin orientation of the final state baryon (the recoil polarization) can be determined. 
Kaon-hyperon production using a spin-polarized photon beam provides access to the beam-, recoil-, target-\footnote{The target polarisation observable can also be accessed by performing a double polarization experiment using a linearly polarised photon beam and measuring the baryon polarisation in the final state.} and to beam-recoil polarization observables. 
The data had a significant impact on the determination of the resonance amplitudes in the mass range above 1.7~GeV. Precision cross section and polarization data,  examples of which are shown in Figure~\ref{KLambda-crs}, span the $K^+\Lambda$ and $K^+\Sigma$ invariant mass range from threshold to 2.9 GeV, hence covering the interesting domain where new states could be discovered. 
Clear resonance-like structures at 1.7~GeV and 1.9~GeV are seen in the $K^+\Lambda$-differential cross section that are particularly prominent and 
well-separated from other structures at backward angles. 
At more forward angles (not shown) t-channel processes become 
prominent and dominate the cross section.
The broad enhancement at 2.2~GeV may also indicate resonant 
behavior although it is less visible at more 
central angles with larger background contributions. 
Similar resonance-like structures are observed in the $K\Sigma$ channel (Figure~\ref{KLambda-crs}(b)).
Examples for different polarisation observables determined for the reaction $\gamma p \to K^+\Lambda$ are shown in the lower row of Figure~\ref{KLambda-crs} for selected bins in the $K^+$-scattering angle in the $\gamma p$ center-of-mass frame. They are compared to 
predictions from ANL-Osaka, BnGa-2014 and to a refit from the BnGa-PWA. The large differences between the curves demonstrate the sensitivity of the data to the underlying dynamics. 
The $K\Lambda$ channel is somewhat easier to understand than the $K\Sigma$ channel, as the 
iso-scalar nature of the $\Lambda$ selects 
isospin-${1/2}$ states to contribute to the 
$K\Lambda$ final state, while both isospin-${1/2}$ and 
isospin-${3/2}$ states can contribute to the 
$K\Sigma$ final state.      
Of course, here, as well as for other final states, only a full partial wave analysis can determine the 
underlying resonances, their masses and spin-parity. Polarization data
are required to disentangle the different amplitudes. 

Energy-dependent analyses have been performed
e.g. at GWU~\cite{Workman:2012jf} as SAID, 
in Mainz as MAID~\cite{A2:2017gwp}, at Kent~\cite{Hunt:2018wqz},  
at JLab~\cite{Julia-Diaz:2007qtz}, 
by the BnGa~\cite{Anisovich:2011fc,CBELSATAPS:2022uad}, 
the J\"ulich-Bonn (J\"uBo) \cite{Ronchen:2022hqk}, the ANL-Osaka~\cite{Kamano:2019gtm} and by other groups. 
A short description of the different methods can be found in Ref.~\cite{Thiel:2022xtb}. 
Here we emphasize that the energy-dependence of a partial-wave amplitude for one particular channel is influenced by other reaction 
channels due to unitarity constraints. To fully describe the energy-dependence 
of a production amplitude, all (or at least the most significant) reaction channels 
must be included in a coupled-channel approach. Many different final states have been measured with high precision off protons 
and partly also off neutrons (bound in a deuteron with a quasi-free proton in the final state).
Polarization data for meson photoproduction off neutrons are, however, still scarce.
A fairly complete list of references can be found in~\cite{Thiel:2022xtb}.
Most data are now included 
in single- and in multi-channel analyses~\footnote{A list of data on photoproduction
reactions including polarization and double-polarization observables can be found at the BnGa web page: https://pwa.hiskp.uni-bonn.de/}. 


%
\begin{table}[pb]
\caption{\label{tab-ut:res} Baryon resonances above the $\Delta$(1232) and below 2300\,MeV
  given in the RPP'2022 in comparison to the resonances considered 
  in the RPP'2010. Resonances with $4*$ in 2010 are not listed here. See text for further discussion.}
\renewcommand{\arraystretch}{1.0}
{\small
\renewcommand{\arraystretch}{1.25}\setlength{\tabcolsep}{+0.35em}
\begin{center}
\begin{tabular}{|c|c|c||c|c|c|}
\hline
 & \footnotesize RPP & \footnotesize RPP  && \footnotesize RPP & \footnotesize RPP  \\[-0.25ex]
 & \footnotesize 2010 & \footnotesize 2022 && \footnotesize 2010 & \footnotesize 2022  \\
\hline
  &&&\\[-3.ex]
  $N(1700)3/2^-$ & ***  & ***      & $\Delta$(1600)3/2$^+$ & *** & ***\bf* \\
  $N(1710)1/2^+$ & ***  & ***\bf*  & $\Delta$(1750)1/2$^+$ & * & * \\
  $N(1860)5/2^+$ &  --  &  \bf**   & $\Delta$(1900)1/2$^-$ & ** & **\bf* \\
  $N(1875)3/2^-$ &  --  &  \bf***  & $\Delta$(1920)3/2$^+$ & *** & ***\\
  $N(1880)1/2^+$ &  --  &  \bf***  & $\Delta$(1930)5/2$^-$ & *** & ***  \\
  $N(1895)1/2^-$ &  --  &  \bf**** & $\Delta$(1940)3/2$^-$ & * & *\bf* \\
  $N(1900)3/2^+$ &  **  &  **\bf** & $\Delta$(2000)5/2$^+$ & ** & ** \\
  $N(1990)7/2^+$ &  **  & **       & $\Delta$(2150)1/2$^-$ & * &  * \\
  $N(2000)5/2^+$ &  **  & **       & $\Delta$(2200)7/2$^-$ & * &  *\bf** \\
  $N(2040)3/2^+$ &  --  & \bf*     &                       &    &\\
  $N(2060)5/2^-$ &  --  & \bf***   &   N(2080)3/2$^-$ & ** & --   \\    
  $N(2100)1/2^+$     &  * & *\bf** & N(2090)$1/2^-$ & *  &     --          \\
  $N(2120)3/2^-$ & --   & \bf***   &  N(2200)5/2$^-$ &  ** & -- \\
\hline
\end{tabular}
\end{center}
}\vspace{-3mm}
\end{table} 
The photoproduction data had a strong impact on the discovery of several new baryon  
states or  
provided new evidence for candidate states 
that had been observed previously but lacked confirmation (e.g.\,\cite{A2:2017gwp,Anisovich:2011fc,Anisovich:2015gia}). Many new decay modes
were discovered, in particular in the photoproduction of $2\pi^0$ and 
$\pi^0\eta$~(\cite{CBELSATAPS:2022uad,CBELSATAPS:2014wvh,CBELSATAPS:2015taz} and references therein).
At the NSTAR'2000 worskhop, 12 $N^*$ and 8 $\Delta^*$ were considered to be established (4*,3*) by the 
Particle Data Group\footnote{In PDG notation: 4* Existence certain, 3* almost certain, 2* evidence
fair, 1* poor}. These numbers increased to 19 $N^*$ and 10 $\Delta^*$ two decades later. Table~\ref{tab-ut:res} lists
the new resonances below 2300\,MeV and those that had not a four-star status in 2010. 
Resonances which had four stars in 2010 are well established and kept their status. These are:  $N(1440)1/2^+$, $N(1520)3/2^-$, \,\,$N(1535)1/2^-$, \,\,$N(1650)1/2^-$, \,\,$N(1675)5/2^-$, 
$N(1680)5/2^+$,  \,\,$N(1720)3/2^+$, \,\,$N(2190)7/2^-$, \,\,$N(2220)9/2^+$, 
$N(2250)9/2^-$, \,\,$\Delta(1620)1/2^-$, \,\,$\Delta(1700)3/2^-$, \,\,$\Delta(1905)5/2^+$, 
$\Delta(1910)1/2^+$, $\Delta(1950)7/2^+$.
A few resonances were removed from the RPP tables. They often had wide-spread mass values,
and the old results were redistributed according to their masses and the new findings. 
Even more impressive is the number of reported decay modes. 
Our knowledge on $N^*$ and $\Delta^*$ decays has at least been doubled.

\section{Regge trajectories}
\begin{figure}[pt]
\resizebox{1.0\columnwidth}{!}{\includegraphics{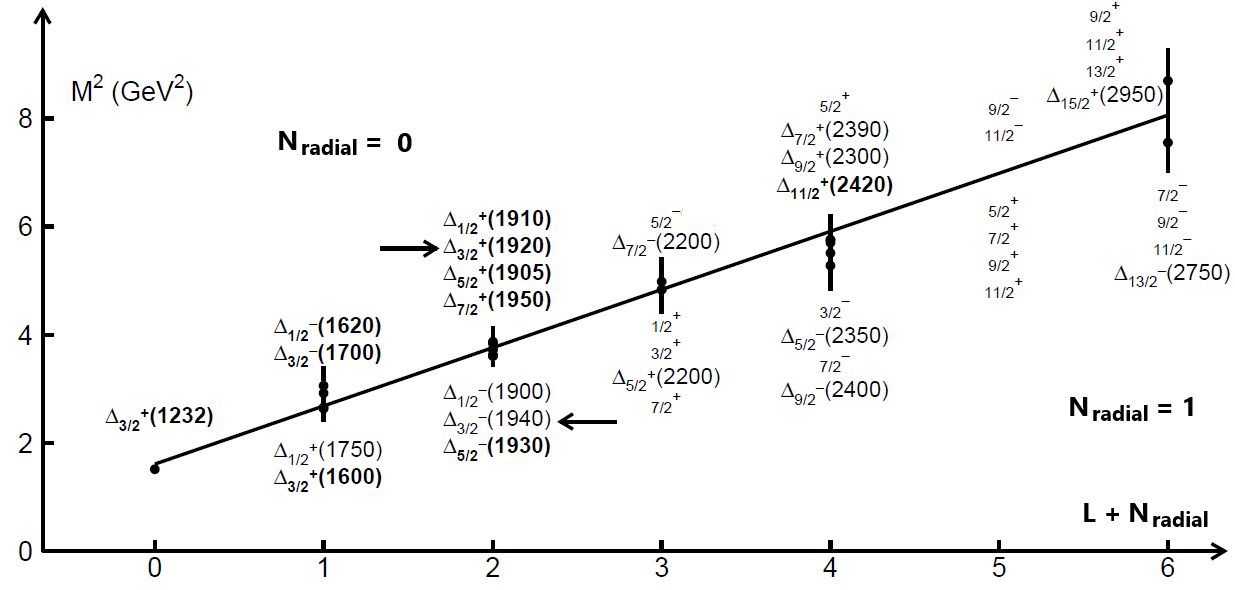}}
\caption{Regge-like trajectory of $\Delta^*$-resonances. Taken from~\cite{Klempt:2008rq}}  
\label{fig:Reggedelta}
\end{figure}

Like mesons, baryons fall onto linear Regge trajectories when their
squared masses are plotted as a function of their total spin $J$ or their intrinsic orbital 
angular momentum $L$. In the case
of $\Delta^*$, the leading trajectory consists of {$\Delta(1232)3/2^+$, 
$\Delta(1950)7/2^+$, \,$\Delta(2420)11/2^+$, \,$\Delta(2950)15/2^+$}. In the
quark model, these have intrinsic orbital angular momenta $L=0,2,4,6$. 
Figure~\ref{fig:Reggedelta} shows the squared $\Delta^*$-masses as a function of $L+N_{\rm radial}$, where $N_{\rm radial}$ indicates the intrinsic radial excitation.
The resonances $\Delta(1910)1/2^+$, $\Delta(1920)3/2^+$, \,$\Delta(1905)5/2^+$ 
\,have \,intrinsic \,$L=2$ \,like $\Delta(1950)7/2^+$, and fit onto the trajectory. Also, there are three
positive-parity resonances that likely have $L=4$ with the $5/2^+$ state missing.
The two $L=1$ resonances $\Delta(1620)1/2^-$ and $\Delta(1700)3/2^-$ 
also have masses close to the linear trajectory. 
Further, there are resonances in which the $\rho$ or $\lambda$ oscillator is excited
radially to $n_\rho=1$ or $n_\lambda=1$ ($N_{\rm radial}=1$). Quark models with a
harmonic oscillator as confining potential predict that resonances belong to
shells. 
Radial excitations are predicted in the
shell $L+2\,N_{\rm radial}$. This is not what we find experimentally: the masses
are approximately proportional to $L+N_{\rm radial}$ if $N_{\rm radial}=1$ is assigned
to $\Delta(1600)3/2^+$, the first radial excitation of $\Delta(1232)3/2^+$, as well as to
the $\Delta(1900)1/2^-$, $\Delta(1940)3/2^-$, $\Delta(1930)5/2^-$ triplet,
to the two members of a partly unseen quartet $\Delta(2350)5/2^-$ and $\Delta(2400)9/2^-$, 
and to $\Delta(2750)13/2^-$ (with $L$=5, $S$=3/2 and $\,N_{\rm radial}$=1). 

Clearly, this is a very simplified picture of the $\Delta^*$ spectrum. The picture is
that of the non-relativistic quark model -- nobody understands why it works\footnote{In addition, we neglect the possible configuration mixing of states in our discussion.}. 
Resonances -- assumed to have the same mass if spin orbit-coupling is neglected-- 
have indeed somewhat different masses. But the gross features of the spectrum of $\Delta^*$ resonances are well reproduced.

The nucleon spectrum is more complicated. First, there are more resonances, and second,
there are two-quark configurations which are antisymmetric in spin and flavor\footnote{These two-quark
configurations are often called {\it good diquarks}. 
They may carry 
orbital-angular momenta, these are not {\it frozen} diquarks.}. 
Due to instanton induced interactions, the relativistic quark model \cite{Loring:2001kx}, expects a lowering of states with the 
respective symmetry. 
Indeed baryons with two-quark configurations which are antisymmetric in spin and flavor ({\it good diquarks}) seem to have 
lower masses than those having {\it bad diquarks}
only. Attempts to include {\it good-diquark}
effects were rather successful~\cite{Klempt:2002vp,Forkel:2008un}. The 
$\chi^2$ for the model-data comparison was twice better for the 2-para\-meter fit than
for quark models~\cite{Klempt:2010du} when the same mass-uncertainties are assumed.  

\section{Hyperons}
Nearly no new data on $\bar K N$ scattering have become available for several decades
except some new data from BNL at very low energy (see Ref.~\cite{Prakhov:2008dc}
and references therein). The reaction
$\gamma p \to K^+\Sigma\pi$ was studied at JLab and helped to understand the
low-energy region \cite{CLAS:2013rjt}.  However,
four groups have re-analyzed $K^-p$ reactions using extensive collections of the old data. 
The new analysis progress was pioneered by the Kent group which performed a 
comprehensive partial wave analysis \cite{Zhang:2013cua,Zhang:2013sva}. 
Energy-independent amplitudes were 
constructed by starting from an energy-dependent fit and by freezing or releasing 
sets of amplitudes. The resulting amplitudes were then fit in a coupled-chan\-nel approach. 
The JPAC group performed coupled-channel fits to the partial waves of the Kent
group. The fit described the Kent
partial waves well while significant discrepancies showed up between data and
the observables calculated from their partial-wave amplitudes~\cite{Fernandez-Ramirez:2015tfa}. 
The ANL-Osaka group used the data set collected by the Kent group and
derived energy-dependent amplitudes based on a phenomenological SU(3) Lagrangian.
Two models were presented which agreed for the leading contributions but which showed
strong deviations for weaker contributions~\cite{Kamano:2014zba,Kamano:2015hxa}.
The BnGa group added further data and tested 
systematically the inclusion of additional states with any set of quantum numbers.
Only small improvements in the fit were found~\cite{Matveev:2019igl,Sarantsev:2019xxm}.

The new studies of old data did not change the situation significantly. Some new decay modes were 
reported, some new but faint signals were found, some were confirmed by one group and missed by others.
Several {\it bumps} were removed from the RPP Tables (for details see~\cite{Klempt:2020bdu}).
As a result, our picture of hyperons (with strangeness $S=-1$) remains unclear. Not even all
states expected in the first $\Lambda$ and $\Sigma$ excitation shell have been seen.
In Table~\ref{tab:sum} all candidates are included.


Very little is known about excited Cascade baryons. A few 
structures in invariant mass spectra
were observed, nearly no spin-parities have been determined. The hope is that at FAIR, JLab
and J-PARC
new $\Xi$'s and $\Omega$'s will be observed and their quantum numbers be determined.
 
\section{QCD expectations}
The spectrum of excited nucleons has been calculated in different approaches. 
We list a few here: QCD on a lattice has been used to calculate
the spectrum of light-quark baryons including hybrid states \cite{Edwards:2011jj}. 
In the Dyson-Schwinger/Bethe-Salpeter approach \cite{Eichmann:2018adq}
the covariant three-body Fadeev-equation is solved in a rainbow-ladder approximation.
The spectra of baryon resonances have been calculated for $J=1/2^\pm$ and
$J=3/2^\pm$, reaching for the $N^*$- and $\Delta^*$-resonances to masses
up to about 2000\,MeV. AdS/QCD \cite{Brodsky:2014yha}
predicts a spectrum of $N^*$ and $\Delta^*$ that is proportional to $L +N_{\rm radial}$. 
Using chiral unitary approaches for the meson-baryon interactions, certain 
baryon resonances can be generated dynamically. 
Various quark models have been developed that treat baryons as bound states of three
quarks with constituent masses, a confinement potential and residual quark-quark interactions.
 At present, 
they are still best suited to discuss
what has been learned from recent results in the spectroscopy of light baryons.

\section{What did we learn within the quark model?}

\subsection{SU(6)$\otimes$O(3) classification}

Table~\ref{tab:sum} lists the observed $N^*$-, $\Delta^*$-,
$\Lambda^*$- and $\Sigma^*$-baryons in a SU(6)$\otimes$O(3) classification.
This classification assumes non-relativistic constituent quarks. It has been
a miracle since the early times of the quark model that this scheme works so
well. But baryon resonances often have a leading  
component in the wave function corresponding to the SU(6)$\otimes$O(3) classification
 even in relativistic calculations.
\begin{table}[pt]
\caption{\label{tab:sum}The spectrum of $N$, $\Delta$, $\Lambda$ and $\Sigma$ excitations. 
The first row shows the quantum numbers of the SU(6)$\otimes$O(3) symmetry group.
$D$ is the dimensionality of the SU(6) group, $L$ the total internal quark orbital
angular momentum, $P$ the parity, $N$ a shell index, $S$ the total 
quark spin, $J$ the total angular momentum. The assignment of particles to 
SU(6)$\otimes$O(3) is an educated guess. 
In the first and second excitation band, all expected states are listed, missing resonances 
are indicated by a $-$ sign. The third band lists only bands
for which at least one candidate exists. 
The states with an index 
are special: above 1700\,MeV, one pair of $\Sigma$ states is expected at about 1750 to 1800\,MeV, 
two pairs at about 2000 to 2050\,MeV. Two pairs marked$^{a}$ are found only. The pairs are shown with 
the three possible assignments. Likewise, $N(2060)$ and $N(2190)$  marked$^{b}$ could form a 
spin-doublet or be members of a spin-quartet. Likely, the observed
pairs of states are mixtures of these allowed configurations (Adapted from \cite{Klempt:2020bdu}).  
}
\includegraphics[width=0.5\textwidth]{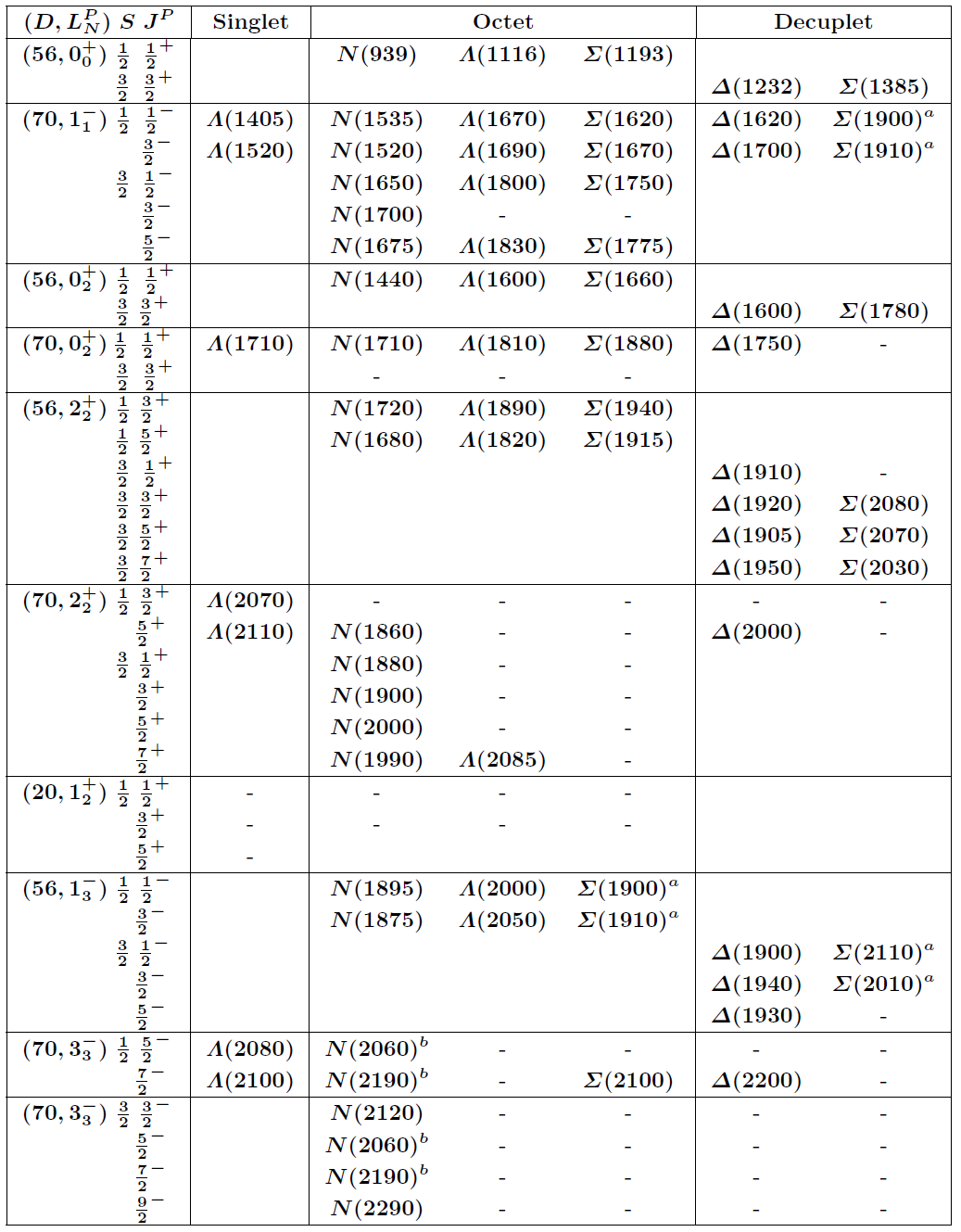}
\end{table}
\begin{figure*}[h!]
\includegraphics[height=0.2\textheight]{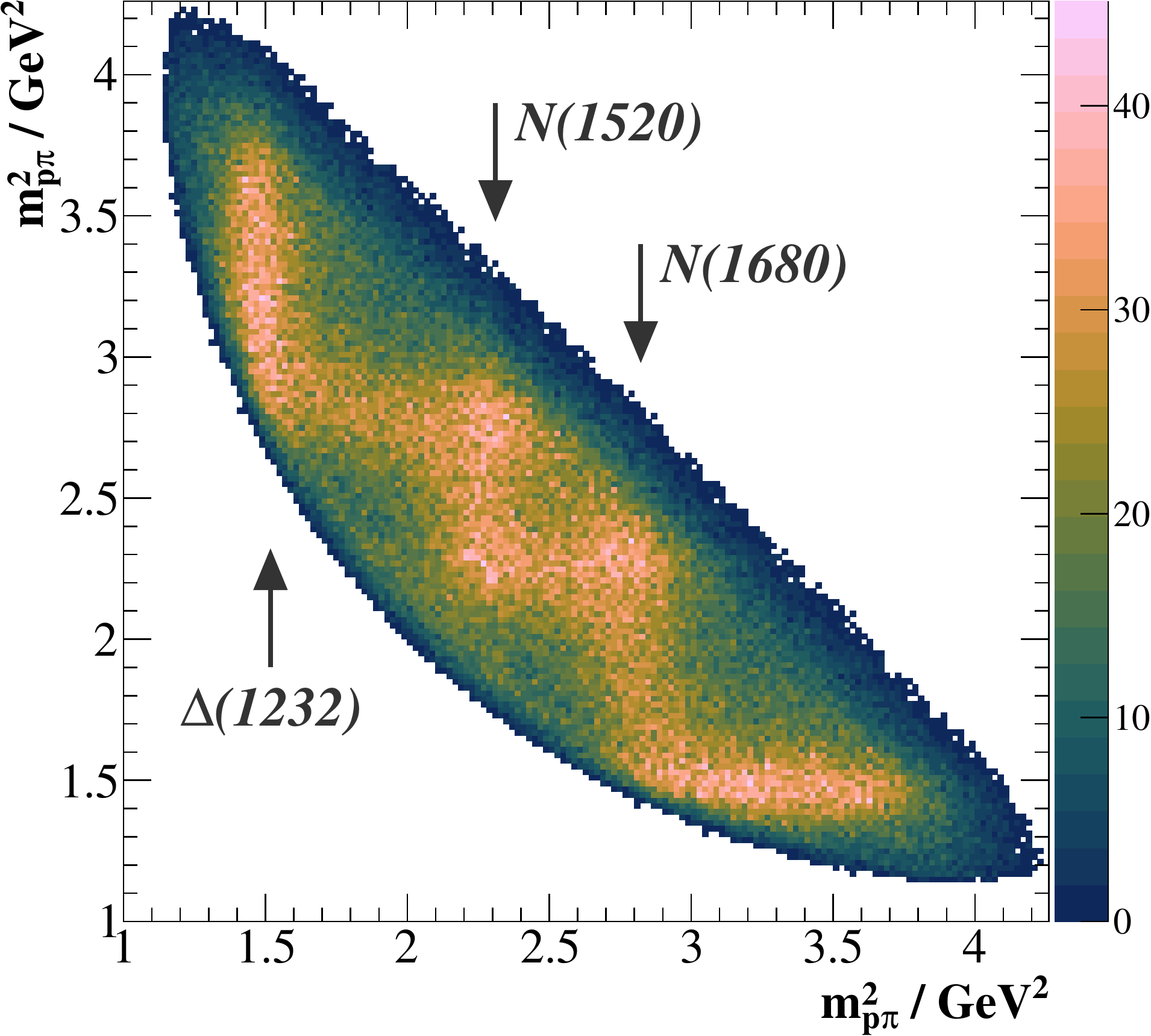} 
\hspace{+0.4cm}
\includegraphics[height=0.22\textheight]{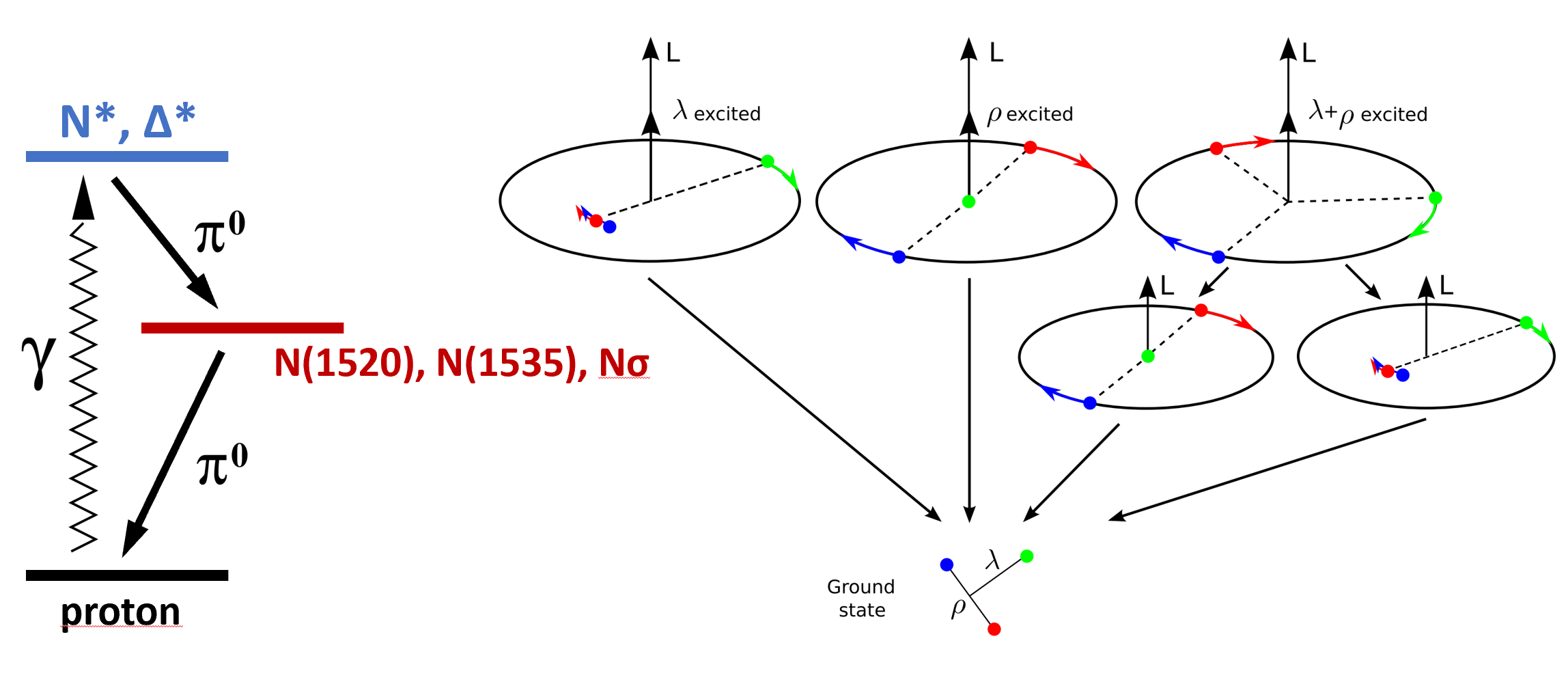}
\caption{\label{fig-ut:cascade-skizze}Left: $\gamma p \to p \pi^0\pi^0$-Dalitz plot for a selected $E_\gamma$-bin of 1900-2100\,MeV (CBELSA/TAPS)~\cite{PMahlberg}, Middle: Cascade decays of resonances via an immediate state. Right: Classical orbits of nucleon excitations with $L=2$ (upper row) and $L=1$ (middle row). Taken from \cite{CBELSATAPS:2015taz}. The first two pictures in the upper row 
show excitations of the $\rho$ and $\lambda$ oscillators, in the third picture both, $\rho$ and $\lambda$ are excited. When both oscillators are excited, de-excitation leads to an excited intermediate state
(middle row).}
\vspace{-3mm}
\end{figure*}
The first excitation shell ($N$$=$$1$) is fairly complete. As expected, there are
five $N^*$'s and two $\Delta^*$'s with negative parity. Of the $\Lambda$ and
$\Sigma$ octet states with negative parity, only the
$J^P=3/2^-$ states are missing\footnote{The $N(1700)3/2^-$ is wider than its 
spin partners and more difficult to identify. This may also be the reason for the
absence of the $J^P=3/2^-$ $\Lambda$ and $\Sigma$ states.}. The two
states $\Lambda(1800)1/2^-$ and $\Sigma(1750)1/2^-$ are interpreted as states
with intrinsic spin 3/2: they seem to be spin partners of 
$\Lambda(1830)5/2^-$ and $\Sigma(1775)5/2^-$.
The doublet of negative-parity decuplet $\Sigma$ states is not uniquely
identified. Expected is this doublet at about 1750\,MeV, and in the (56,1$_3^-$)-configuration a second
doublet at about 2050\,MeV and, finally, a triplet at about the same mass. The
analysis found (poor) evidence for two doublets, marked $^{a}$ in Table~\ref{tab:sum}.
The singlet states $\Lambda(1405)1/2^-$  and $\Lambda(1520)3/2^-$
deserve a more detailed discussion.

At higher masses, some choices are a bit arbitrary: Because of its mass,
$N(1900)3/2^+$ belongs to the second excitation shell. It may have 
intrinsic quark spin 1/2 or 3/2, both with $L=2$. Further, there should be
a $3/2^+$ radially excited state with $L=0$. These three states can mix. Only 
one of the states is clearly identified. In any case,
quark models predict three resonances with $J^P=3/2^+$ in this mass range while only one is found. Also missing is a doublet of states with $L=1$ belonging to the 20plet
in SU(6)$\otimes$O(3).\footnote{The RPP lists three more $N^*/\Delta^*$-resonances: $N(2040)3/2^+$, $\Delta (2150)1/2^-$, which need confirmation and $N(2100)1/2^+$ which we assign to the 4th shell.} The production of this doublet is expected 
to be strongly suppressed for reasons to be discussed below.

Only few hyperons are known that can be assigned to the second excitation shell.
The interpretation of some $\Lambda$ resonances as SU(3) singlet configuration
is plausible but not at all compelling. 

\subsection{Missing resonances}

In the spectrum of $N^*$ and $\Delta^*$, the first excitation shell
is complete, in the second shell, 21 states are expected (two of them
likely not observable in $\pi N$-elastic scattering or in single/double meson photoproduction), 16 are seen,
three are missing. To a large extend, the {\it missing-resonance} problem
is solved for $N^*$ and $\Delta^*$: there are no frozen diquarks. Admittedly, five of the 
resonances are not yet ``established", i.e.
have not (yet?) a 3* or 4* status. 

In the third shell, only few resonances are known, but the number
of expected resonances is quite large and the analysis challenging: 
45 $N^*$ and $\Delta^*$, likely
with widths often exceeding 300\,MeV, are expected
to populate an about 400\,MeV wide mass range. 
\begin{table}[ph]
\begin{scriptsize}
\renewcommand{\arraystretch}{1.3}
\begin{tabular}{|c|c|c|c|}
\hline
        &1. shell & 2. shell&3. shell\\[-0.5ex]
 $J^P$\hspace{-1mm}&\hspace{-1mm}$1/2,\ 3/2,\ 5/2^-$\hspace{-1mm}&\hspace{-1mm}$1/2\ 3/2\ 5/2\ 7/2^+$\hspace{-1mm}&\hspace{-1mm} \ $1/2^- - 9/2^-$ \\
\hspace{-1mm}Masses\hspace{-2mm}& 1500 - 1750&1700 - 2100& 1900 - 2300\\
				\hline
$N$     & 5:\quad  2  \  \  2 \  \ 1    & 13:\quad  4 \  \ 5 \  \ 3 \  \  1   & 30:\quad 7 \,9  \,8 \,5  \,1 \\  
$\Delta$&  2:\quad  1  \  \  1 \  \ -    & \ 8:\quad  2 \  \ 3 \  \ 2 \  \  1   & 15:\quad 3  \,5 \,4  \,2 \,1 \\
\hline
\end{tabular}
\end{scriptsize}
\end{table}
\subsection{Three-quark dynamics in cascade decays}

The CBELSA/TAPS collaboration studied cascade decays of high mass resonances via an
intermediate resonance down to the ground state nucleon. 
The analyses were based on a large data 
base of photoproduction data including final states such as
$\gamma p \to p\pi^0\pi^0$ and $p\pi^0\eta$ (see~\cite{CBELSATAPS:2022uad,CBELSATAPS:2014wvh} and Refs. therein). The Dalitz plot of Fig.~\ref{fig-ut:cascade-skizze}, shows very 
clearly band-like structures due to the occurrence of baryon resonances in the intermediate state. 
It was observed that the positive parity $N^*$- and $\Delta^*$-resonances at a mass of about 1900\,MeV show a very different 
decay pattern. The four $N^*$-resonances $N(1880)1/2^+$, $N(1900)3/2^+$, $N(2000)5/2^+$, $N(1990)7/2^+$
decay with an average branching fraction of $(34\pm 6)\%$ into  $N\pi$ and $\Delta\pi$ and with a 
branching fraction of $(21\pm 5)\%$ into the orbitally excited states $N(1520)3/2^-\pi$, $N(1535)1/2^-\pi$, and $N\sigma$. 
The four $\Delta^*$-states,
$\Delta(1910)1/2^+$, $\Delta(1920)3/2^+$, $\Delta(1905)5/2^+$ and $\Delta(1950)7/2^+$ have
an average decay branching fraction into $N\pi /\Delta\pi$ of $(44\pm 7)\%$ while their
branching fraction into the excited states mentioned above is almost negligible,
only $(5\pm 2)\%$~\cite{CBELSATAPS:2022uad}.  
At the first sight, this is very surprising. 

The difference can be traced to the different wave functions. 
The spin and the flavor wave functions of the four
$\Delta^*$-states are both symmetric with respect to the exchange of any two quarks, the
spatial wave function needs to be symmetric as well. 
This means that - having a three-quark-picture in mind - that either
the $\rho$- or the $\lambda$-oscillator is excited to $\ell=2$, the
other one is not excited. (There is a mixture of the two possibilities 
$\ell_{\rho}=2$, $\ell_{\lambda}=0$ or $\ell_{\lambda}=2$, $\ell_{\rho}=0$). 
If this state decays, the orbital angular momentum is carried away and
the decay products are found preferentially in their ground state.

The four $N^*$-states have a spatial wave function with mixed symmetry. Thus 
the spatial wave function has one part which is mixed-symmetric  and one part
which is mixed anti-symmetric. In the latter one, both oscillators are excited
simultaneously ($\ell_{\rho}=\ell_{\lambda}=1$). If this state decays, 
one of the excitations remains in the decay product as illustrated 
in Fig.~\ref{fig-ut:cascade-skizze}. A similar
argument has been used by Hey and Kelly~\cite{Hey:1982aj} to explain
why the 20'plet in the second excitation
shell of Fig.~\ref{tab:sum} cannot be formed in a $\pi N$ scattering experiment.
For the 20'plet the spacial wave function is entirely antisymmetric, both oscillators 
are excited simultaneously, and there is no other component in the wave function. 
A single-step excitation is suppressed.

\subsection{Parity doublets?}
\begin{figure}
\centering
\begin{tabular}{cc}
  \hspace*{-0.4cm}
  \includegraphics[width=0.25\textwidth]{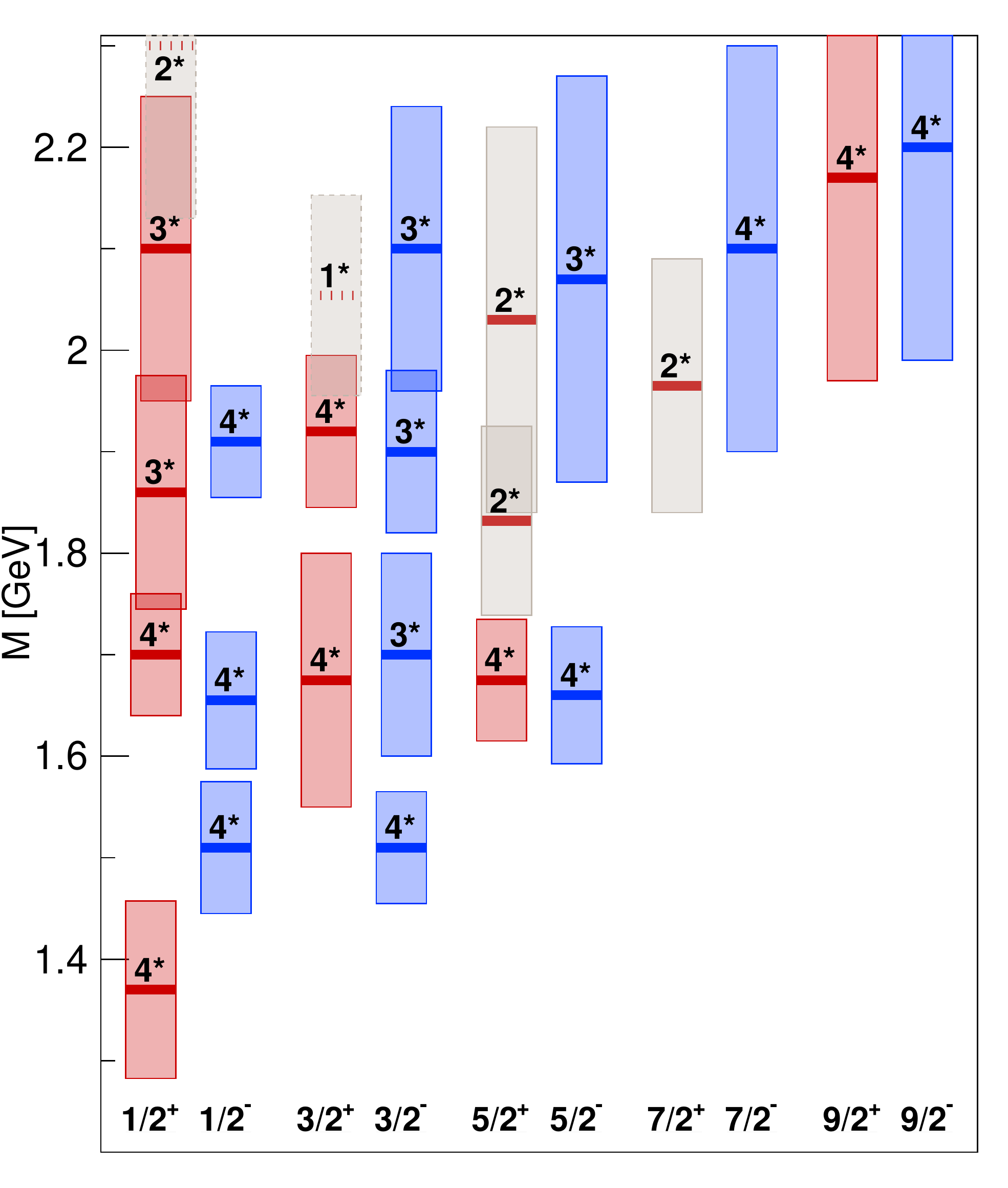} 
  & \hspace*{-0.6cm}
  \includegraphics[width=0.25\textwidth]{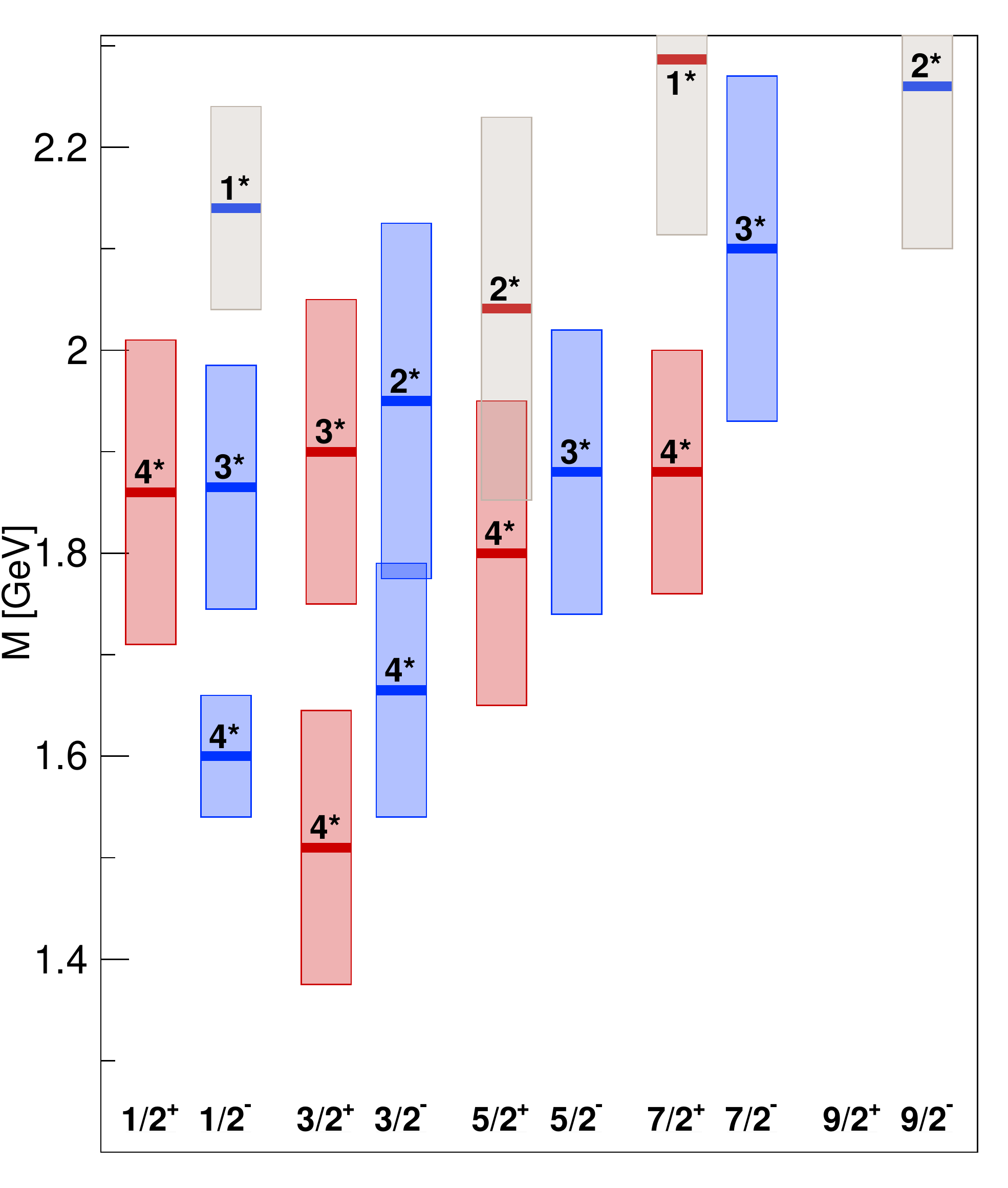} 
\end{tabular}
\caption{\label{fig-ut:parity} N$^*$- (left) and $\Delta^*$-resonances (right) above $\Delta(1232)$
  for different spin and parities $J^p$.
  For each resonance, the real part of the pole position $Re(M_R)$ is given together with
  a box of length $\pm Im(M_R)$, using the PDG estimates.
  $2\cdot Im(M_R)$ corresponds to the total width of the resonance.
  RPP star ratings are also indicated. If no pole positions are given in the RPP
  (above the line), the RPP Breit-Wigner estimates for masses and widths
  are used instead. This is indicated by dashed resonance-mass lines and dashed lines
  surrounding the boxes. If no RPP-estimates are given, the values above the line
  have been averaged and the states are shown as gray boxes. This may indicate
  one measurement above the line only. $\Delta$(1750)$1/2^+$ is not included,
as there is no RPP-value given above the line.\vspace{-2mm} }
\end{figure}

The spontaneous breaking of the chiral symmetry leads to the large mass gap observed
between chiral partners: the masses of the $\rho(770)$ meson with spin-parity
$J^P=1^-$ and its chiral partner $a_1(1260)$ with $J^P=1^+$ differ by about 500\,MeV,
those of the $J^P=1/2^+$ nucleon and $N(1535)1/2^-$ by about 600\,MeV. In contrast 
to  quark-models expectations and lattice QCD calculations~\cite{Edwards:2011jj} 
higher-mass baryons are often observed in parity doublets
(see Fig.~\ref{fig-ut:parity}), in pairs of resonances having about the same mass,
the same total spin $J$ and opposite parities.
\begin{figure}[pt]
\begin{tabular}{cc}
\hspace{-2mm}\includegraphics[width=0.42\columnwidth,height=1.005\columnwidth]{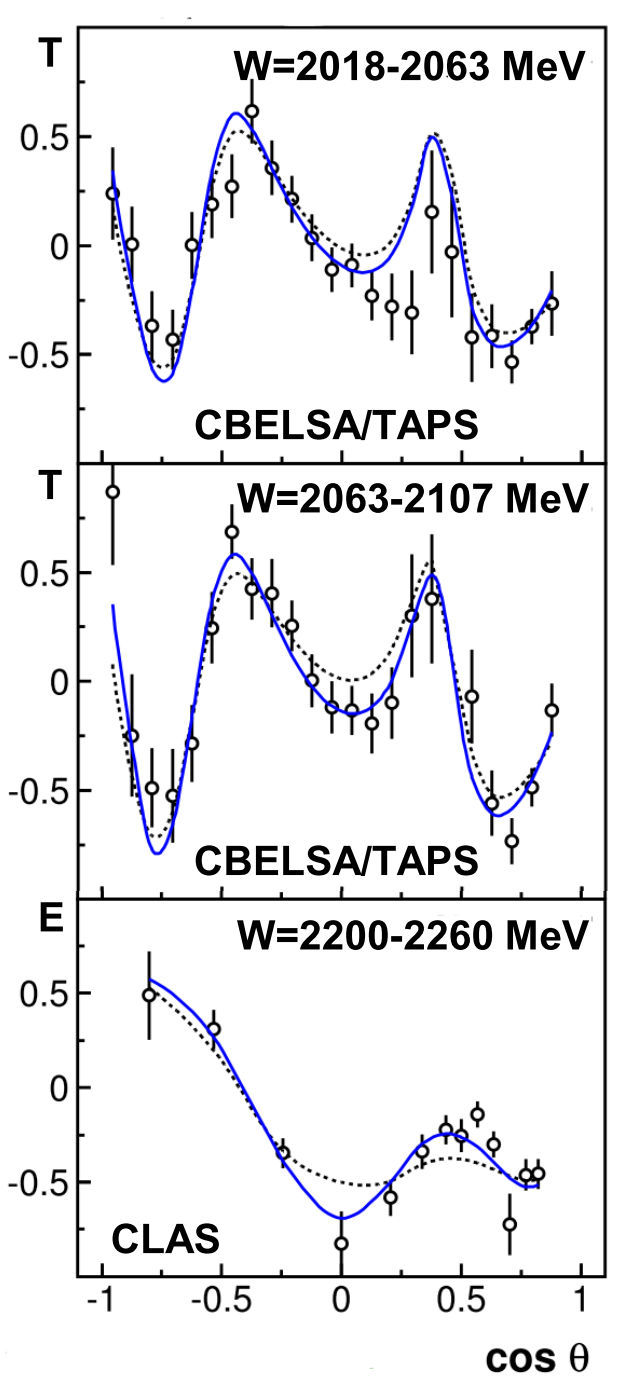}
\hspace{-2mm}\includegraphics[width=0.58\columnwidth,height=\columnwidth]{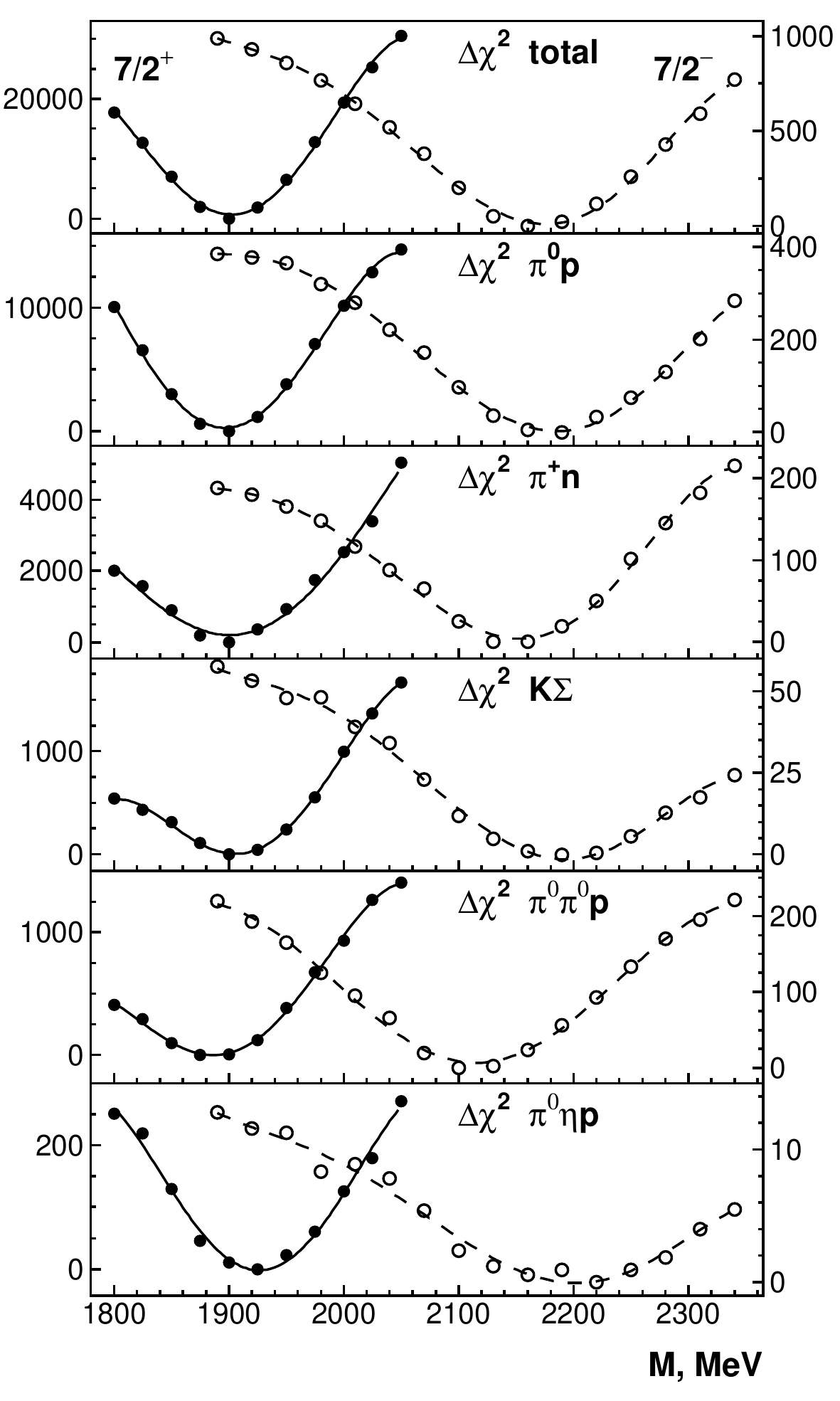}
\end{tabular}
\vspace{-2mm}
\caption{\label{fig-ut:parity-chi2}Left: The new polarization observables $T$ and $E$ shown for selected 
mass bins (see~\cite{Anisovich:2015gia} for refs. to the data). The fit curves represent the best fits with (solid) and without (dashed) inclusion
of $\Delta(2200)7/2^-$. Right: The increase in pseudo-$\chi^2$ of the
  fit to a large body of pion- and photo-produced reactions when the mass of
  $\Delta(1950)7/2^+$ (solid points) or $\Delta(2200)7/2^-$ (open circles) is scanned.
  The scale on the left (right) abscissa refers to the $7/2^+$ ($7/2^-$) partial wave.
  The curves are to guide the eye. Adapted/taken from~\cite{Anisovich:2015gia}. \vspace{-2mm}
}
\end{figure}
This observation and similar observations in meson spectrum has led to the suggestion
that chiral symmetry might be effectively restored in highly excited
hadrons~\cite{Glozman:1999tk,Glozman:2003bt}. Then, all high-mass resonances
should have a parity partner. This is a testable prediction.

In the mass region of 1900\,MeV a quartet of well known
positive parity $\Delta^*$ states exists, consisting of 
$\Delta(1910)1/2^+$, $\Delta(1920)3/2^+$, $\Delta(1905)5/2^+$, and $\Delta(1950)7/2^+$. Figure~\ref{fig-ut:parity} shows the parity partners 
of the first three states: $\Delta(1900)$ $1/2^-$, $\Delta(1940)3/2^-$, and $\Delta(1930)5/2^-$. 
However, the four-star $\Delta(1950)7/2^+$ has no close-by $\Delta(xxx)7/2^-$-state
that could serve as parity partner. Where is the closest  $\Delta^*$ with
$J^P$ = $7/2^-$\,? 
Figure~\ref{fig-ut:parity-chi2} shows a resonance scan over the mass region of interest~\cite{Anisovich:2015gia}. There is clear evidence for
$\Delta(2200)7/2^-$ (which was upgraded from 1$^*$ to 3$^*$ based on  this result).
But its mass difference to $\Delta(1950)$\-$7/2^+$ is too large. These
two states are no parity partners!

Within the quark model and the SU(6)$\otimes$O(3)-syste\-matics, the four positive-parity 
$\Delta^*$'s  have $L=2,S=3/2$ that couple to $J^P=1/2^+$,$\cdots$, $7/2^+$. 
The natural assignment for the three negative-parity $\Delta^*$'s is that they
form a triplet with $L=1$ and $S=3/2$. Then, they must have one unit of radial excitation.
The four positive-parity $\Delta$-states belong to the
$2\hbar\omega$ shell and the negative-parity states to the $3\hbar\omega$ shell.
With masses considered to be proportional to $L+N_{\rm radial}$, these seven states are expected to have about the same mass.
$\Delta(2200)7/2^-$ has $L=3, S=1/2$ and its expected mass is higher. 
We note that
$\Delta(2400)9/2^-$ has $L=3, S=3/2$, and we assume $N_{\rm radial}=1$ for this state
(as well as for $\Delta(2750)13/2^-$, see Fig.~\ref{fig:Reggedelta}).

\section{Dynamically generated resonances}
\subsection{$N^*$'s and $\Delta^*$'s}
Apart from $\Lambda(1405)1/2^-$ that will be discussed below, the first
dynamically generated resonance was the nega\-tive-parity 
$N(1535)1/2^-$~\cite{Kaiser:1995cy}. At the 1995 International Conference on the 
Structure of Baryons, Santa Fe, New Mexico, there was a heated discussion
between Weise, defending his new approach, and Isgur who
argued that $N(1535)1/2^-$ is well understood within the quark model and no new
approach is needed. For some time, there was even the idea that there could be
two overlapping states but this is excluded by data. 
{Later, in~\cite{Bruns:2010sv,Mai:2012wy}
$N(1535)1/2^-$ and $N(1650)1/2^-$, were both shown to be generated dynamically}
but not $\Delta(1620)1/2^-$\footnote{It should be mentioned that not only the SU(6)$\otimes$O(3)-systematics in the spectrum seems to indicate a 3-quark-nature of $N(1535)1/2^-$ and $N(1650)1/2^-$ but also the electroproduction results discussed in the following section~\ref{Burkert} indicate that $N(1535)1/2^-$ is a 3-quark state with little meson-baryon contribution only ($Q^2$ dependence of the transition form factor $A_{1/2}$).}. 
An important question remains: Are (qqq)-resonance poles and 
dynamically generated poles different
descriptions of the same object or do they present different (orthogonal) states?

\subsection{The $\Lambda(1405)1/2^-$}
The $\Lambda(1405)1/2^-$ mass is very close to the $N\bar K$ threshold. 
Kaiser, Waas and Weise~\cite{Kaiser:1996js} proved that the resonance can be 
generated dynamically from $N\bar K-\Sigma\pi$ coupled-channel
dynamics. Oller and Meissner~\cite{Oller:2000fj} studied the $S$-wave $N\bar K$
interactions in a relativistic chiral unitary approach based
on a chiral Lagrangian obtained from the interaction of the octet of pseudoscalar 
mesons and the ground state baryon octet and found two isoscalar resonances 
in the $\Lambda(1405)1/2^-$ mass region and one isovector state. In a subsequent paper~\cite{Jido:2003cb},
Jido {\it et al.} studied the the effects of SU(3) breaking on the results in detail.
These two papers had an immense impact on the further development. It
is the only result in light-baryon spectroscopy that is in clear
contradiction to the quark model. It introduces a new state $\Lambda(1380)1/2^-$,
that has no role in a quark model, it enforces an interpretation of 
$\Lambda(1405)1/2^-$ as mainly SU(3) octet resonance,  
and it interprets $\Lambda(1670)1/2^-$ as high-mass partner of $\Lambda(1405)1/2^-$. The $\Lambda(1405)1/2^-$ and $\Lambda(1670)1/2^-$ would then be the strange partners of the $N(1535)1/2^-$ \,and \,the \,$N(1650)1/2^-$. 
\,In \,quark \,models, $\Lambda(1405)1/2^-$ is a mainly SU(3) singlet resonance and the octet states 
$\Lambda(1670)1/2^-$ and $\Lambda(1800)1/2^-$ are the strange partners
of $N(1535)1/2^-$ and $N(1650)1/2^-$ (see Table~\ref{tab:sum}). In \,the \,quark-model \,interpretation, \,the 
\,hyperon \,states $\Lambda(1405)1/2^-$ and $\Lambda(1670)1/2^-$ have close-by $J^P=3/2^-$ 
partners (the $J^P=3/2^-$-partner of $\Lambda(1800)1/2^-$ is missing but there is $\Lambda(1830)5/2^-$). 
The masses of the mainly octet states are about
130\,MeV above their non-strange partners. 

This conflict initiated an attempt to fit (nearly) all existing data relevant for
$\Lambda(1405)1/2^-$ in the BnGa approach~\cite{Anisovich:2020lec}. 
The data could be fit with
one single resonance in the $\Lambda(1405)1/2^-$ region but were also 
compatible, with a slightly worsened $\chi^2$, with a description using
two resonances with properties as obtained in the chiral unitary approach.

\section{Outlook}
There is not yet a unified picture of baryons. 
Regge-like trajectories ($M^2\propto L+N_{\rm radial}$) are best described
by AdS/QCD. Unitary effective field theories describe consistently meson-baryon interactions
and some resonances are generated dynamically from their interaction. The quark model is useful
to understand cascade decays of highly excited states and is indispensable to discuss
the full spectrum including missing resonances.The symmetry of quark pairs, symmetric or anti-sym\-metric
with respect to their exchange, 
has a significant impact on baryon masses. They could be due to effective gluon exchange.
More likely seems an interpretation by quark and gluon condensates, e.g. by 
instanton-induced interactions. 
Based on the new high quality (polarized) photoproduction data, new baryon resonances
were discovered and our knowledge of properties of existing resonances has increased considerably. 
Yet, our understanding 
is still unsatisfactory mirroring the complexity of QCD in the non-perturbative 
regime. New results from lattice QCD are eagerly awaited and new experiments are needed to understand the spectrum and the properties of baryon resonances in further detail. Those include further precise photoproduction experiments measuring polarisation observables not only off the proton but also off the neutron as well as multi-meson final states. Strange baryon resonances need to be addressed. Other production processes such as electroproduction, $\bar{p}p$-anni\-hilation, experiments with $\pi$- or $K$-beams and baryon resonances produced in $J/\psi$ or $\psi^\prime$-decays will also  contribute to improve our understanding of the bound states of the strong interaction. \\

\bibliography{lights}
\bibliographystyle{ieeetr}
\end{document}